\documentclass[preprint,aps]{revtex4}

\newcommand {\bea}{\begin{eqnarray}}
\newcommand {\eea}{\end{eqnarray}}
\newcommand {\be}{\begin{equation}}
\newcommand {\ee}{\end{equation}}
\newcommand {\qslash}{q\!\!\!/}
\newcommand {\muslash}{\mu\!\!\!/}

\begin{document}

\preprint{TRI-PP-99-30}

\title{Patterns of Symmetry Breaking in QCD at High Baryon Density}

\author{Thomas Sch\"afer}

\affiliation{TRIUMF, 4004 Wesbrook Mall\\
Vancouver, BC, Canada, V6T2A3}

\begin{abstract}
  We study the structure of QCD at very large baryon density
for an arbitrary number of flavors $N_f$. We provide evidence
that for any number of flavors larger than $N_f=2$ chiral
symmetry remains broken at asymptotically large chemical
potential. For $N_c=N_f=3$, chiral symmetry breaking 
follows the standard pattern $SU(3)_L\times SU(3)_R\to
SU(3)$, but for $N_f>3$ unusual patterns emerge. We study
the case $N_f=3$ in more detail and calculate the magnitude
of the chiral order parameters $\langle\bar\psi\psi\rangle$
and $\langle(\bar\psi\psi)^2\rangle$ in perturbative QCD. 
We show that, asymptotically, $\langle\bar\psi\psi\rangle^{1/3}$
is much smaller than $\langle(\bar\psi\psi)^2\rangle^{1/6}$.
The result can be understood in terms of an approximate discrete 
symmetry.  
\end{abstract}
\maketitle

\section{Introduction}

  The phase structure of matter at non-zero baryon density
has recently attracted a great deal of interest. In particular,
it has been emphasized that quark matter at very high density
is expected to behave as a color superconductor \cite{ARW_98,RSSV_98}.
The behavior of hadronic matter in this regime is of interest in 
understanding the structure of compact astrophysical objects 
and the physics of heavy ion collisions in the regime of 
maximum baryon density.  Moreover, it was realized that matter 
at very high density exhibits many non-perturbative phenomena, 
such as a mass gap and chiral symmetry breaking, in a regime where 
the coupling is weak \cite{ARW_98b,SW_98b}. This means that some of 
the ``hard'' problems of QCD can be studied in a systematic fashion. 

  At very high density the natural degrees of freedom are
quasiparticles and holes in the vicinity of the Fermi
surface. Since the Fermi momentum is large, asymptotic freedom
implies that the interaction between quasiparticles is weak.
In QCD, because of the presence of unscreened long range gauge 
forces, this is not quite true. Nevertheless, we believe 
that this fact does not essentially modify the argument
\cite{Son_98}. However, as we know from the theory of 
superconductivity the Fermi surface is unstable in the presence 
of even an arbitrarily weak attractive interaction. At very large
density, the attraction is provided by one-gluon exchange between 
quarks in a color anti-symmetric $\bar 3$ state. QCD at high density 
is therefore expected to behave as a color superconductor 
\cite{Frau_78,Bar_77,BL_84}.

  Color superconductivity is characterized by the breakdown 
of color gauge invariance. As usual, this statement has to 
be interpreted with care. Local gauge invariance cannot really
be broken \cite{Eli_75}. Nevertheless, spontaneously broken
gauge invariance is a useful concept. We can fix the gauge,
introduce a gauge non-invariant order parameter and study 
its effect on gauge invariant correlation functions. The 
most important gauge invariant consequence of superconductivity
is the appearance of a mass gap, through the Meissner-Anderson-Higgs 
phenomenon. The formation of a mass gap is of course also characteristic
of a confined phase. Indeed, it is known that in general 
Higgs and confined phases are continuously connected \cite{FS_79}.
 
  Color superconductivity may also lead to the spontaneous
breaking of global symmetries. It is this phenomenon that
we wish to study in more detail. Broken global symmetries
lead to the appearance of Goldstone bosons, and determine
the low energy effective description of the system. It
is one of the remarkable properties of the color superconducting
phase in $N_f=3$ QCD that the pattern of broken global 
symmetries exactly matches that of QCD at low density. 
In this work, we would like to establish the global symmetry 
breaking pattern in QCD with $N_f=2,3,\ldots,6$. In 
particular, we wish to show that the result in the case
$N_f=3$ is generic in the sense that for any number of 
flavors larger than 2, chiral symmetry is broken but a
vector-like flavor symmetry remains. We also wish to
study the most interesting case, QCD with $N_f=3$, in 
more detail. We calculate the magnitude of the totally
symmetric and anti-symmetric order parameters, and the 
magnitude of the chiral condensates $\langle\bar\psi\psi
\rangle$ and $\langle(\bar\psi\psi)^2\rangle$. We comment
on a number of unusual features of chiral symmetry breaking
in QCD at large density.

\section{The effective potential}
\label{sec_veff}

  In this section we wish to introduce a simple energy 
functional that captures the essential dynamics of QCD at 
high baryon density. We will use this functional in the 
following section in order to analyze the ground state
of QCD with an arbitrary number of flavors. 

  Our construction is guided by the renormalization group 
analysis of \cite{SW_98,EHS_98}. In this work we classified
possible instabilities of the Fermi surface, and assessed 
their relative importance, for small and short-range, but 
otherwise arbitrary couplings near the Fermi surface. It was 
found that the dominant instability corresponds to scalar 
diquark condensation. The analysis does not fix the color 
and flavor channel of this instability uniquely, because
there are two equally enhanced interactions. One gluon 
exchange, which dominates for weak coupling, is attractive 
in the color anti-symmetric $\bar 3$ channel, and 
favors one of these interactions. The same is also true
for instanton induced interactions, that are likely to 
play an important role at moderate densities. As the 
couplings evolve towards the Fermi surface, the attractive
interaction in the color $\bar 3$ channel will grow, 
while the repulsive interaction in the color symmetric 6 
channel is suppressed. The dominant coupling is a
color and flavor anti-symmetric interaction of the 
form
\bea
\label{laa}
 {\cal L} &=& G
 \big(\delta^{ac}\delta^{bd}-\delta^{ad}\delta^{bc}\big)
 \big(\delta_{ik}\delta_{jl}-\delta_{il}\delta_{jk}\big)
 \left\{ \left(\psi^a_i C\gamma_5 \psi^b_j \right)
     \left(\bar\psi^c_k C\gamma_5 \bar\psi^d_l \right)
     - \left( C\gamma_5\leftrightarrow C\right )\right\},
\eea
where $a,b,\ldots$ are color indices and $i,j\ldots$ 
are flavor indices. In the following, we shall use the 
notation ${\cal K}^{abcd}_{ijkl}$ for the color-flavor structure
of the interaction. 

 In full QCD, there are unscreened magnetic gluon exchanges
and the interaction between quarks is not short range. This
problem was first studied in \cite{Son_98}. We expect that
this effect does not modify our results for the structure
of the ground state, but only the magnitude of the gap. We 
check this explicitly in the case of $N_f=3$ in section
\ref{sec_cfl}. In this section, we also study the effect
of color-symmetric interactions. It was also pointed out
that for asymptotically large chemical potential, the gap
equation in higher partial wave channels becomes degenerate
with the $s$-wave gap equation \cite{Son_98}. Subleading 
effects are expected to lift this degeneracy. Using the 
methods developed in \cite{SW_99b} we have checked 
that the $s$-wave gap is
bigger than higher partial wave gaps. We therefore 
expect the ground state to be an $s$-wave superconductor.
The only exception is QCD with only one flavor, since 
in this case a color $\bar 3$ order parameter cannot 
have spin zero. This situation is of physical interest
for the behavior of real QCD with two light and one intermediate 
mass flavor. For $m_s$ larger than some critical value, QCD
has a phase with separate pairing in the $ud$ and
$s$ sectors \cite{SW_99,ABR_99}. In this work, we will 
not consider the phase structure of $N_f=1$ QCD.

  In order to determine the structure of the ground state
we have to calculate the grand canonical potential of the system 
for different trial states. Since the interaction is
attractive in s-wave states, it seems clear that the 
dominant order parameter is an s-wave, too. We then 
only have to study the color-flavor structure of the 
primary condensate. We assume that the condensate 
takes the form
\bea
\label{order}
 \langle \psi^a_i C\gamma_5\psi^b_j\rangle
 = \phi^{ab}_{ij}.
\eea
$\phi^{ab}_{ij}$ is a $N_f\times N_f$ matrix in flavor
space and a $N_c\times N_c$ matrix in color space. The
Pauli principle requires that $\phi^{ab}_{ij}$ is symmetric 
under the combined exchange $(ai)\leftrightarrow (bj)$. 

  We calculate the effective potential using the bosonization
method. For this purpose, we introduce collective fields
$\Delta^{ab}_{ij}$ and $\bar\Delta^{ab}_{ij}$ with the 
symmetries of the order parameter (\ref{order}). We add to
the fermionic action a term $(4G)^{-1}{\cal K}^{abcd}_{ijkl}
\Delta^{ab}_{ij}\bar\Delta^{cd}_{kl}$ and integrate over the 
dummy variables $\Delta^{ab}_{ij}$ and $\bar\Delta^{ab}_{ij}$. 
We then shift the integration variables to eliminate the 
interaction term (\ref{laa}). So far, no approximations have
been made. We now assume that the collective fields are 
slowly varying, and that $\Delta^{ab}_{ij}$ can be replaced
by its vacuum expectation value. In this case, we can perform 
the integration over the fermionic fields and determine the 
grand canonical potential as a function of the gap matrix 
$\Delta^{ab}_{ij}$. 

 The integration over the fermions is performed using the 
Nambu-Gorkov formalism. We introduce a two component field 
$\Psi=(\psi,\bar\psi^T)$. The inverse quark propagator takes 
the form
\be
\label{sinv}
S^{-1}(q) = \left(\begin{array}{cc}
 \qslash+\muslash-m &  {\cal K}\cdot\bar\Delta \\
 {\cal K}\cdot\Delta  & (\qslash-\muslash+m)^T 
\end{array}\right).
\ee
The grand canonical potential is now given by
\be 
\Omega(\Delta) = \frac{1}{2}{\rm Tr}\left[\log(S_0^{-1}S)\right]
 + \frac{1}{4G}\Delta\cdot{\cal K}\cdot\bar\Delta.
\ee
In order to evaluate the logarithm, we have to diagonalize 
the mass matrix ${\cal M}={\cal K}\cdot \Delta$. Let us denote 
the corresponding eigenvalues by $\Delta_\rho\,(\rho=1,\ldots,
N_cN_f)$. These are the physical gaps of the $N_fN_c$ fermion 
species. If we neglect the quark masses $m$, the grand potential is 
\be
\label{gcp}
 \Omega(\Delta) = \sum_{\rho} \left\{ 
 -\int\frac{d^3p}{(2\pi)^3}\left(
 \sqrt{(p-\mu)^2+\Delta_\rho^2}+\sqrt{(p+\mu)^2+\Delta_\rho^2}\right)
 + \frac{1}{G}\Delta_\rho^2\right\}
\ee
The momentum integral has an ultraviolet divergence. This integral
can be regularized by introducing a cutoff $\Lambda$ or, more 
generally, by including a form factor $F(p)$. In this case it
would seem that the properties of the grand potential depend
on a number of parameters, such as the coupling constant $G$,
the chemical potential $\mu$ and the cutoff $\Lambda$. In the 
weak coupling limit, however, the grand potential should only
depend on the value of the gap on the Fermi surface, and not
on the exact momentum dependence of the interaction. This can
be made manifest by introducing a renormalized grand potential.
Following the work of Weinberg \cite{Wei_94}, we have
\be
\label{renp}
 \Omega_{ren}(\Delta) = \sum_{\rho}\left\{ 
 \frac{\mu^2}{4\pi^2}\Delta_\rho^2
  \left(\log\left(\frac{\Delta_\rho}{\xi}\right)-1\right)
 + \frac{1}{G(\xi)}\Delta_\rho^2 \right\}.
\ee
Here, $\xi$ is a renormalization scale. The grand potential
is independent of $\xi$, since the scale dependence of the
first term is canceled by the scale dependence of the coupling 
constant $G$. The coupling constant satisfies the renormalization
group equation discussed in \cite{EHS_98,SW_98}. 

 The grand potential (\ref{renp}) depends on $N_c(N_c-1)
N_f(N_f-1)/4$ parameters. We minimize this function numerically.
After we determine the matrix $\Delta^{ab}_{ij}$ that minimizes
the grand potential we study the corresponding symmetry breaking
pattern. Without superconductivity, there are $N_f^2-1$ global 
flavor symmetries for both left and right handed fermions, as 
well as $N_c^2-1$ local gauge symmetries. Superconductivity 
reduces the amount of symmetry. In order to find the residual
symmetry group we study the second variation of the order 
parameter $\delta^2\Delta/(\delta\theta_i\delta\theta_j)$, where
$\theta_i\;(i=1,\ldots,N_f^2+N_c^2-2)$ parameterizes the flavor and
color transformations. Zero eigenvalues of this matrix correspond
to unbroken color-flavor symmetries. 

\section{Color superconductivity in QCD with $N_c=3$
colors and $N_f$ flavors}
\label{sec_nf}

 The results of our numerical study for QCD with three colors 
and $N_f=2,\ldots,6$ flavors are summarized in Table 1. In
the following, we will study each of these cases in more detail.

  The two flavor case is special. In this case, the order
parameter
\be
\label{order_2}
 \Delta^{ab}_{ij} = \Delta \epsilon^{3ab}\epsilon_{ij}
\ee
only breaks color $SU(3)\to SU(2)$. The chiral $SU(2)_L
\times SU(2)_R$ symmetry remains unbroken. At this level,
the Fermi surfaces of the up and down quarks of the third
color remain intact. A careful study of the quantum numbers
of the low energy states shows that chiral symmetry is realized
in terms of massless protons and neutrons \cite{SW_99}. The 
proton and neutron are composites of the quark and Higgs fields. 
Subleading interactions can generate a gap for these states. 
The exact nature of this gap is hard to determine, even in the 
limit of very large chemical potential. 

  For $N_f$ larger than 2, the gauge symmetry is always 
completely broken, and all quarks acquire a gap. In addition
to that, we find that the preferred order parameter always
involves a coupling between the color and flavor degrees
of freedom. This means that the original flavor symmetry
is broken, but some vector-like symmetry which is a combination 
of the original flavor and color symmetries remains. 

  In the case of three flavors we find that the preferred 
order parameter is of the form
\be
\label{order_3}
\Delta^{ab}_{ij} = 
 \Delta (\delta_i^a\delta_j^b-\delta_j^a\delta_i^b).
\ee
This is the color-flavor locked phase suggested in 
\cite{ARW_98b}. Both color and flavor symmetry are completely 
broken. There are eight combinations of color and flavor symmetries 
that generate unbroken global symmetries. The symmetry breaking
pattern is 
\be
\label{sym_3}
SU(3)_L\times SU(3)_R\times U(1)_V\to SU(3)_V .
\ee
This is exactly the same symmetry breaking that QCD exhibits 
at low density. This has led to the idea that in QCD with 
three flavors, the low and high density phases might be 
continuously connected \cite{SW_98b}. We also note that the 
quark mass gaps fall into representations ([8]+[1]) of the 
unbroken vector symmetry. Note that in the present analysis, 
which only takes into account the leading interaction, the
order parameter is completely anti-symmetric in both color 
and flavor. In the case $N_f=N_c$, however, there is a
more general order parameter $\Delta^{ab}_{ij}=\Delta_1
\delta^a_i\delta^b_j +\Delta_2\delta^a_j\delta^b_i$ which 
has both symmetric and anti-symmetric components and has 
the same residual symmetry. We will discuss this situation
in more detail in the next section.

 In the case $N_f=4$ the numerical results indicate that 
the most favorable order parameter is given by
\be
\label{order_4}
\Delta^{ab}_{ij} =  \Delta \epsilon^{abc}\eta^c_{ij}
 = \Delta \epsilon^{abc} (\epsilon_{ijc}+\delta_{ic}\delta_{j4}
  - \delta_{jc}\delta_{i4} ),
\ee
where $\eta^a_{ij}$ is the 't Hooft symbol. There is a second,
degenerate, solution where we replace $\eta^a_{ij}\to\bar
\eta^a_{ij}=\epsilon_{ija}-\delta_{ia}\delta_{j4}+ \delta_{ja}
\delta_{i4}$. The 't Hooft symbol describes the isomorphism
$SU(2)\times SU(2)\equiv O(4)$. Consider the $O(4)$ generators
$(M^{\rho\sigma})_{\mu\nu}=i(\delta^\rho_\mu\delta^\sigma_\nu-
\delta^\sigma_\mu\delta^\rho_\nu)$. The $O(4)$ generators can
be decomposed into $SU(2)_L\times SU(2)_R$ generators
$\eta^a_{\mu\nu}(M^{\mu\nu})_{\rho\sigma}$ and $\bar\eta^a_{\mu\nu}
(M^{\mu\nu})_{\rho\sigma}$. Using this result one can show
that the $N_f=4$ order parameter (\ref{order_4}) realizes
the symmetry breaking pattern 
\be
\label{sym_4}
 SU(4)_L\times SU(4)_R\times U(1)_V\to SU(2)_V\times SU(2)_V .
\ee
The generators of the $SU(2)\times SU(2)$ symmetry are
\be
\left(\epsilon^{abc}N_{bc}+
    \frac{1}{2}\eta^a_{\mu\nu}M^{\mu\nu}\right), \hspace{0.5cm}
\left(\epsilon^{abc}N_{bc}+
    \frac{1}{2}\bar\eta^a_{\mu\nu}M^{\mu\nu}\right).
\ee
Here, $N_{bc}$ are $O(3)\subset SU(3)$ color generators and 
$M_{\mu\nu}$ are $O(4)\subset SU(4)$ flavor generators.
We note that for $N_f=4$, and indeed for any $N_f$ larger than
three, we cannot realize the same symmetry breaking that we 
have at low density. This means that even though chiral symmetry
remains broken at very large density, there has to be a phase
transition that separates the high and low density phases.
  
 In the case $N_f=5$ we were unable to find a compact expression
for the energetically favored order parameter. The numerical
results indicate that $N_f=5$ QCD realizes the symmetry 
breaking pattern
\be
\label{sym_5}
SU(5)_L\times SU(5)_R\times U(1)_V \to SU(2)_V.
\ee
The residual symmetry is smaller than for any other number of 
flavors. This is also reflected in the fact that the condensation
energy is comparably small. 

  If $N_f$ is a multiple of $N_c$, the dominant gap corresponds 
to multiple embeddings of the $N_f=N_c$ order parameter. For $N_f=6$, 
we have
\be
\label{order_6}
\Delta^{ab}_{ij} = \Delta \epsilon^{abc} 
 (\epsilon_{ijc}+\epsilon_{(i-3)(j-3)c}).
\ee
This order parameter corresponds to the symmetry breaking pattern
\be
\label{sym_6}
SU(6)_L\times SU(6)_R\times U(1)_V \to SU(3)_V\times U(1)_V
 \times U(1)_A.
\ee
The $SU(3)$ symmetry is a double embedding of the $SU(3)$ 
that appear in the $N_f=3$ color-flavor locked phase. The 
original $U(1)_V$ and approximate $U(1)_A$ symmetries
are broken, but a new $U(1)_V\times U(1)_A$ appears. This symmetry is a 
subgroup of the original $SU(6)_L\times SU(6)_R$ flavor 
symmetry which rotates the two $3\times 3$ blocks in equation 
(\ref{order_6}). The extra $U(1)$ symmetries may be 
broken by higher order condensates.

  In this section we have analyzed the color-flavor structure
of the superconducting ground state using a BCS-like energy
functional. Another method, which has proven to be very useful 
in the context of liquid $^3He$ and other systems \cite{he3}, 
is the Landau-Ginzburg functional. This method was applied 
to color superconductivity in \cite{BL_84,PR_98}. In this
case we construct the most general energy functional that 
is consistent with the symmetries of QCD and a simple 
polynomial in the order parameter. In four dimensions, it
is usually sufficient to keep terms that are at most quartic 
in the fields. We can analyze the possible ground states
by studying the minima of the Landau-Ginzburg functional. 

  There is one restriction that one has to keep in mind. 
At $T=0$ the free energy of the system in not an analytic
function of the gap so that, strictly speaking, the free energy 
cannot be expanded as a power series in the order parameter. Only 
in the vicinity of the finite temperature phase transition
does the expansion in powers of the order parameter have a 
firm foundation. 

 Nevertheless, the Landau-Ginzburg description is very useful
in describing a wealth of phenomena, even at $T=0$. If we 
restrict ourselves to color and flavor anti-symmetric 
order parameters we can represent the order parameter 
matrix $\Delta^{ab}_{ij}$ by the field $\phi^a_i$, where
$a$ is an index in the anti-symmetric $[N_c(N_c-1)/2]$
of color and $i$ in the $[N_f(N_f-1)/2]$ of flavor. 
Color and flavor invariance imply that the Landau-Ginzburg
effective potential has the form
\be
\label{vlg}
V = -m^2{\rm tr}\big(\phi^\dagger\phi\big)
  + \lambda_1 \Big[ {\rm tr}\big(\phi^\dagger\phi\big)\Big]^2
  + \lambda_2 {\rm tr}\Big[ \big(\phi^\dagger\phi\big)^2\Big],
\ee
where $(\phi^\dagger\phi)^{ab}=(\phi^a_i)^*\phi^b_i$, and $m^2,
\lambda_1$ and $\lambda_2$ are arbitrary parameters. Of course, we 
can equally well write (\ref{vlg}) in terms of $(\phi\phi^\dagger)_{ij}
=\phi_i^a(\phi_j^a)^*$. Note that there are no $\det(\phi)$
terms, because such a term would violate the $U(1)$ of
baryon number. The effective potential only depends on 
the eigenvalues of $\phi^\dagger\phi$. This means that the effective
potential only depends on a number of parameters, $N_c(N_c-1)/2$ 
(or $N_f(N_f-1)/2$, whichever is smaller). 

 For $N_f=2$ there is only one quartic term and the minimum
occurs for $\phi^a=\Delta\delta^{a3}$, as we would expect. In 
the case $N_f=3$ there are two possible minima, $\phi_i^a=
\Delta\delta^a_i$ and $\phi_i^a=\Delta\delta^{a3}\delta_{i3}$. 
These minima correspond to the color-flavor locked phase 
and the $N_f=2$ phase. The groundstate depends on the values
of the coupling constants. If $\lambda_2>0$, the ground state
is the color-flavor locked phase. Extending the mean field
description described in the last section to $T\neq 0$, we 
can check that indeed $\lambda_1=0$ and $\lambda_2>0$, see
also \cite{BL_84,PR_98}. 

 For $N_f>3$ the effective potential does not depend on
the number of flavors. This means that there are a large
number of ground states all of which are degenerate with
an embedding of the $N_f=3$ color-flavor locked phase. The 
ground states of the BCS functional discussed above are
minima of the Landau-Ginzburg potential, but the effective
potential (\ref{vlg}) does not distinguish partially gapped 
from fully gapped states. In order to construct an effective 
energy functional for which the fully gapped state is the 
true minimum one has to include higher powers in the order 
parameter. This, of course, would also introduce additional
parameters and we will not pursue this problem here.

\section{More on color-flavor-locking in QCD with
$N_c=N_f=3$}
\label{sec_cfl}

 In this section we wish to examine the color-flavor-locked
state in QCD with three colors and flavors in somewhat more 
detail. For this purpose, we will concentrate on the regime
of very large chemical potential, $\mu\gg\Lambda_{QCD}$, in
which perturbative calculations are possible. The determination
of the gap in perturbative QCD has recently attracted some
attention \cite{Son_98,SW_99b,PR_99b,HMSW_99,BLR_99,HS_99}. The 
main conclusion is that the gap is dominated by almost collinear
magnetic gluon exchanges. The magnitude of the gap is 
\be
\label{gap_oge}
\Delta_0 \simeq 256\pi^4(2/N_f)^{5/2}\mu g^{-5}
   \exp\left(-\frac{3\pi^2}{\sqrt{2}g}\right).
\ee
We should emphasize that, strictly speaking, this result
contains only an estimate of the preexponent. This estimate
is obtained by collecting the leading logarithms from 
electric and magnetic gluon exchanges \cite{SW_99b}. There
are corrections of order $O(1)$ that originate from improved
matching of the gap function at $p_0\simeq\Delta_0$ and
$p_0\simeq g\mu$, a self-consistent treatment of the
Meissner effect, and, possibly, higher order perturbative
corrections. Nevertheless, the main point is that (\ref{gap_oge}) 
is the result of a well-defined calculation that can be 
systematically improved. 

 In previous works, the gap equation was always studied
for a one-component color-flavor anti-symmetric order
parameter. This is appropriate for QCD with two flavors,
but in the case of more than two flavors the gap equation
is more complicated. In the previous section we studied
a more general $N_cN_f\times N_cN_f$ dimensional
gap matrix but restricted ourselves to short range, 
color and flavor antisymmetric interactions. The 
restriction to antisymmetric gap matrices is probably
justified for $N_f\neq N_c$, but for $N_f=N_c$ the 
symmetric and anti-symmetric order parameters have 
the same global symmetry, so there is no symmetry
reason for the symmetric gap parameter to be zero. 

  In three flavor QCD, the order parameter has the form
\be
\label{nf3_cfl}
\Delta^{ab}_{ij} = 
 \Delta_A (\delta_i^a\delta_j^b-\delta_j^a\delta_i^b)
+\Delta_S (\delta_i^a\delta_j^b+\delta_j^a\delta_i^b).
\ee
We can now repeat the perturbative calculation of the 
gap using this particular ansatz for the order parameter. We 
will follow the method described in \cite{SW_99b}. Just 
as in the $N_f=2$ case, the gap depends on frequency
and the Dirac structure of the gap matrix is proportional
to $C\gamma_5(1+\vec\alpha\cdot\hat p)/2$. The only
new ingredient is that we have to take into account the 
more complicated color-flavor structure when we calculate
the Nambu-Gorkov propagator. As in the last section, this 
is most easily accomplished by viewing (\ref{nf3_cfl}) as 
a matrix in a $N_cN_f$ dimensional color-flavor space. 
The eigenvalues of this matrix are 
\be
\label{cfl_evals} 
\Delta_8 = \Delta_A-\Delta_S,
\hspace{1cm}
\Delta_1 = 2\Delta_A+4\Delta_S ,
\ee
where the subscript indicates the degeneracy.
For three degenerate flavors the normal components of 
the inverse Nambu-Gorkov propagator are proportional to the 
unit matrix in color-flavor space, so they remain diagonal as 
the anomalous components are diagonalized. We can now 
determine the propagator by inverting the Nambu-Gorkov
and Dirac structure as in the $N_f=2$ case. The off-diagonal
propagator $S_{21}$ needed in the gap equation is a diagonal 
matrix in color-flavor space with entries
\be
\label{S21}
 S^i_{21}(q) = \frac{1}{2}(C\gamma_5)
 \frac{\Delta_i(1-\vec\alpha\cdot\vec q)}{q_0^2-(q-\mu)^2-\Delta_i^2},
\ee 
with $\Delta_i=(\Delta_8,\Delta_8,\ldots,\Delta_1)$ as in 
(\ref{cfl_evals}). Having determined the propagator we have
to calculate the color factor. It is given by
\be
\label{c_a}
c_A = \frac{1}{4} (\lambda^A)^T_{ab}
  (\delta_i^a\delta_j^b- \delta_i^b\delta_j^a)
 (\lambda^A)_{cd} 
  = -\frac{N_c+1}{2N_c}
 (\delta_i^c\delta_j^d- \delta_i^d\delta_j^c)
\ee
for the color-flavor anti-symmetric gap and
\be
\label{c_s}
c_S = \frac{1}{4} (\lambda^A)^T_{ab}
  \left(\delta_i^a\delta_j^b+ \delta_i^b\delta_j^a \right)
 (\lambda^A)_{cd} 
  = \frac{N_c-1}{2N_c}
 \left(\delta_i^c\delta_j^d+ \delta_i^d\delta_j^c \right)
\ee
for the symmetric gap. The anti-symmetric color factor $c_A$
agrees with the result for the $N_f=2$ order parameter. We can now
project the gap equation on the color-flavor anti-symmetric and 
symmetric structures. We find
\bea
\label{eliash1}
\Delta_A(p_0) &=& \frac{g^2}{18\pi^2} \int dq_0
 \log\left(\frac{b\mu}{|p_0-q_0|}\right) \left\{
 \frac{2}{3}\frac{\Delta_A(q_0)-\Delta_S(q_0)}
         {\sqrt{q_0^2+(\Delta_A(q_0)-\Delta_S(q_0))^2}}
 \right.\nonumber \\
& & \hspace{5cm}\left.\mbox{}
 +\frac{1}{6}\frac{2\Delta_A(q_0)+4\Delta_S(q_0)}
         {\sqrt{q_0^2+(2\Delta_A(q_0)+4\Delta_S(q_0))^2}}
 \right\},\\
\Delta_S(p_0) &=& \frac{g^2}{18\pi^2} \int dq_0
 \log\left(\frac{b\mu}{|p_0-q_0|}\right) \left\{
  \frac{1}{6}\frac{\Delta_A(q_0)-\Delta_S(q_0)}
         {\sqrt{q_0^2+(\Delta_A(q_0)-\Delta_S(q_0))^2}}
 \right.\nonumber \\
& & \hspace{5cm}\left.\mbox{}
 -\frac{1}{12}\frac{2\Delta_A(q_0)+4\Delta_S(q_0)}
         {\sqrt{q_0^2+(2\Delta_A(q_0)+4\Delta_S(q_0))^2}}
 \right\},
\eea
where $b=256\pi^4(2/N_f)^{5/2}g^{-5}$. This is a complicated 
system of coupled equations, but the situation simplifies in
the weak coupling limit. In this case, we can assume that
$\Delta_S\ll\Delta_A$. The equation for $\Delta_A$ then 
becomes
\be
\label{eliash2}
\Delta_A(p_0) = \frac{g^2}{18\pi^2} \int dq_0
 \log\left(\frac{b\mu}{|p_0-q_0|}\right) \left\{
 \frac{2}{3}\frac{\Delta_A(q_0)}
         {\sqrt{q_0^2+\Delta_A(q_0)^2}}
 +\frac{1}{3}\frac{\Delta_A(q_0)}
         {\sqrt{q_0^2+(2\Delta_A(q_0))^2}}
 \right\}.
\ee
If it were not for the factor 2 in the denominator of
the second term in the curly brackets, this would be 
exactly the same equation as the one we found for the
simple $N_f=2$ order parameter. We can take the factor
2 into account in an approximate way by rescaling the 
integration variable in the second term. In this way
we can reduce equation (\ref{eliash2}) to the gap
equation for the $N_f=2$ order parameter with the 
coefficient $b$ rescaled by a factor $2^{-1/3}$. 
This means that
\be
\label{dela_cfl}
 \Delta_A\simeq 2^{-1/3} 256\pi^4(2/N_f)^{5/2}\mu g^{-5}
   \exp\left(-\frac{3\pi^2}{\sqrt{2}g}\right).
\ee
The equation for $\Delta_S$ can be analyzed in a 
similar fashion. We find
\be
\label{dels_cfl}
 \Delta_S \simeq \frac{g}{\pi}\frac{\sqrt{2}\log(2)}{36}\Delta_A .
\ee
This implies that, formally, $\Delta_S$ is suppressed by
one power of the coupling constant, $g$. In addition to that,
we note that the numerical coefficient in (\ref{dels_cfl}) is
quite small, so that $\Delta_S\ll\Delta_A$ even if $g$ is 
not small. 

 These results are easily generalized to an arbitrary number of 
flavors and colors with $N_f=N_c=N$. The eigenvalues of the 
gap matrix are
\be
 \Delta_{N^2-1} = \Delta_A-\Delta_S ,
\hspace{1cm}
 \Delta_1 = (N-1)\Delta_A + (N+1)\Delta_S, 
\ee
and the $N_c$ dependence of the anti-symmetric and symmetric
color factor is given in (\ref{c_a},\ref{c_s}). From these
results we find that the color-flavor-locked gap is given
by
\bea
\label{cfl_n}
\Delta_A &=&   256\pi^4(N-1)^{-\frac{N-1}{2N}}
(2/N)^{5/2}\mu g^{-5}
   \exp\left(-\frac{\pi^2}{g}\sqrt{\frac{6N}{N+1}}\right),\\
\Delta_S &=& \frac{g}{6\pi}
   \left(\frac{N-1}{2N}\right)^2
   \left(\frac{6N}{N+1}\right)^{1/2}
   \log(N-1) \Delta_A.
\eea
The origin of the various factors of $N$ in (\ref{cfl_n}) is
easily explained. The factor in the exponent is the color
factor that comes from the tree level scattering amplitude
of two quarks in a color anti-symmetric state. The factor
$N^{-5/2}$ originates from the flavor dependence
of the screening mass, and the factor $(N-1)$ raised to 
the power $-(N-1)/(2N)$ comes from the structure of the 
color-flavor locked state. This factor implies that for 
large $N$, the CFL gap is suppressed by a factor $\sqrt{N}$ 
with respect to the gap in the $N_f=2$ phase. As a consequence,
for $N>3$ the color-flavor locked state (\ref{nf3_cfl}) is not 
necessarily the energetically favored state. If, on the other
hand, we take the large $N_c$ limit with $N_f=3$ fixed, we
find multiple embeddings of the $N_c=N_f=3$ state. If the 
large $N_c$ limit is taken with the conventional scaling 
$g^2N_c=const$, the superconducting gap is suppressed by 
$\exp(-\sqrt{N_c})$. This means that for very large $N_c$
and $N_f$ fixed, the superconducting ground state is disfavored 
compared to a chiral density wave \cite{DGR_92,SS_99}.

  In addition to the gap equation, we would also like to
study the condensation energy. To order $g^2$ the grand 
potential can be calculated from \cite{FM_77,BL_84}
\be
\label{f_ng}
\Omega = \frac{1}{2}\int\frac{d^4q}{(2\pi)^4}
 \left\{-{\rm tr}\left[S(q)\Sigma(q)\right]
       +{\rm tr}\log\left[S_0^{-1}(q)S(q)\right]\right\},
\ee
where $S(q)$ and $\Sigma(q)$ are the Nambu-Gorkov 
propagator and proper self energy given in \cite{SW_99b}. 
The grand potential has the form $\Omega=8f(\Delta_8)+f(\Delta_1)$
where $\Delta_{8,1}$ are the singlet and octet gaps. The
functional $f(\Delta)$ is given by
\be
f(\Delta) = \frac{\mu^2}{\pi^2}  \int dp_0 \left\{ 
-\frac{\Delta(p_0)}{\sqrt{p_0^2+\Delta(p_0)^2}}
+\sqrt{p_0^2+\Delta(p_0)^2}-p_0 \right\}.
\ee
This result is very similar to the result in the case 
of short range interactions \cite{BL_84}. The only difference
is that the energy dependence of the gap function acts as an 
explicit cutoff in the integrals. We can calculate the integrals
using the approximate solution of the Eliashberg equation 
derived by Son \cite{Son_98}. Numerically we find that the 
condensation energy scales to very good accuracy as
\be
f(\Delta) = \frac{\mu^2}{4\pi^2} \Delta_0^2
 \log\left(\frac{\Delta_0}{\mu}\right),
\ee
where $\Delta_0=\Delta(p_0=0)$. This is very similar to the 
result we found for short range interactions, equation (\ref{renp}).
This suggests that the results obtained in section \ref{sec_nf} 
remain valid in the more general case of long range interactions. 
In particular, we can calculate the energy gain of the color-flavor
locked state over the $N_f=2$ state. The result is only very weakly
dependent on the gap in a very wide range of $\Delta/\mu$. In 
the weak coupling limit, we find $\epsilon(CFL)/\epsilon(N_f\!=\!2)
\simeq 1.9$. This ratio is somewhat smaller than what one would expect 
based on the number of gaps, $9/4\simeq 2.2$.

\section{Chiral symmetry breaking}
\label{sec_csb}

 The most interesting aspect of the color-flavor locked
phase is that chiral symmetry is broken, and that the 
form of the corresponding low energy effective action 
agrees with QCD at low density \cite{SW_98b,CG_99}. But while 
the coefficients of the effective lagrangian are complicated, 
non-perturbative quantities in QCD at low density, they can 
be calculated perturbatively at high density. 

  In this section we would like to begin this program
by calculating the chiral order parameter in the color-flavor 
locked phase. As a first step, we have to calculate the superfluid 
condensate. For a single fermion species we have
\be
\phi=\langle \psi C\gamma_5
      \frac{1}{2}(1+\vec\alpha\cdot\hat{p})\psi\rangle
    = \frac{\mu^2}{\pi^2}\int dp_0
      \frac{\Delta(p_0)}{\sqrt{p_0^2+\Delta(p_0)^2}}.
\ee
In the weak coupling limit, the integrand can be written
as a total derivative using the differential equation for 
$\Delta(p_0)$ \cite{Son_98,MSW_99}. The result is
\be
\label{phi}
  \phi = 2\left(\frac{\mu^2}{2\pi^2}\right)
       \frac{3\sqrt{2}\pi}{g}\Delta .
\ee
In the CFL state, the color-flavor structure of the 
condensate is more complicated. Just like the gap
matrix, the condensate can be written as
\be
\label{phi_cfl}
\langle\psi_i^a C\gamma_5
   \frac{1}{2}(1+\vec\alpha\cdot\hat{p})\psi_j^b\rangle
 = \phi_A (\delta_i^a\delta_j^b-\delta_j^a\delta_i^b)
  +\phi_S (\delta_i^a\delta_j^b+\delta_j^a\delta_i^b).
\ee
The two condensates $\phi_{A,S}$ can be determined
from the octet and singlet gap parameters. Not 
surprisingly, we find that $\phi_S$ is small in the 
weak coupling limit. In the case of $\phi_A$, all 
color and flavor factors drop out and
\be
\phi_A = 2\left(\frac{\mu^2}{2\pi^2}\right)
       \frac{3\sqrt{2}\pi}{g}\Delta .
\ee
The chiral structure of the superfluid order 
parameter is $\psi C\gamma_5\psi=\psi_R\psi_R-
\psi_L\psi_L$. This means that pairing takes 
place between quarks of the same chirality.
A convolution of two superfluid order 
parameters $(\psi_R\psi_R)(\bar\psi_L\bar\psi_L)$
will then directly yield the gauge invariant 
order parameter $\langle(\bar\psi_L\psi_R)
(\bar\psi_L\psi_R)\rangle$. In the weak 
coupling limit, factorization is valid and
we find
\be
\langle (\bar\psi_L\psi_R)(\bar\psi_L\psi_R)\rangle
 = \frac{1}{4}
   \left( 12\phi_A^2+24\phi_S^2 \right)
 \simeq 3\phi_A^2 . 
\ee
In the same way we can calculate many other 
four fermion operators, like  
\be
\langle (\bar\psi\psi)^2\rangle = 
   \langle (\bar\psi_L\psi_R)(\bar\psi_L\psi_R)\rangle
 + \langle (\bar\psi_R\psi_L)(\bar\psi_R\psi_L)\rangle
= 6\phi_A^2.
\ee
Here, we have used $\langle(\bar\psi_L\psi_R)
(\bar\psi_R\psi_L)\rangle\simeq 0$. This is a consequence
of the chiral structure of the superfluid order parameter
mentioned above. 

  There is an important point that we need to emphasize here. The 
vacuum expectation value $\langle(\bar\psi\psi)^2\rangle$ is not
an order parameter for chiral symmetry breaking.
This is most obvious in the case of two flavors. In this case,
$(\bar\psi\psi,\bar\psi i\gamma_5\vec\tau\psi)$ transforms as 
a 4-vector under the chiral $SU(2)_L\times SU(2)_R=O(4)$.
Chiral symmetry restoration then implies $3\langle(\bar\psi\psi)^2
\rangle=\langle(\bar\psi i\vec\tau\gamma_5\psi)^2\rangle$, not 
$\langle(\bar\psi\psi)^2\rangle=0$. In the case of $N_f=3$, 
a true order parameter for chiral symmetry breaking is
given by \cite{KKS_98}
\be
\label{4f_o1}
{\cal O}_1 = (\bar\psi_L\gamma_\mu\lambda^a\psi_L)
             (\bar\psi_R\gamma_\mu\lambda^a\psi_R),
\ee
where $\lambda^a$ is a flavor generator. 
This operator transforms as $(8,8)$ under $SU(3)_L\times
SU(3)_R$, so a non-zero expectation value of ${\cal O}_1$
definitely implies that chiral symmetry is broken. Fierz
rearranging ${\cal O}_1$, we get an alternative order 
parameter 
\be
\label{4f_o2}
{\cal O}_2 = \frac{2(N_f^2-1)}{N_f^2}
       (\bar\psi_L\psi_R)(\bar\psi_R\psi_L)
 -\frac{1}{N_f}
       (\bar\psi_L\lambda^a\psi_R)
       (\bar\psi_R\lambda^a\psi_L).
\ee
We can calculate the vacuum expectation values of ${\cal O}_{1,2}$
in the mean field approximation. Both turn out to be zero. This 
does not, of course, imply that chiral symmetry is unbroken. The
fact that $\langle {\cal O}_{1,2}\rangle$ vanishes is due to 
an accidental symmetry of the mean field approximation. The 
superfluid order parameter is invariant under $(Z_2)_L\times 
(Z_2)_R$ symmetries that act on the fermion and anti-fermion 
fields separately. Because ${\cal O}_{1,2}$ do not have this
symmetry, they cannot acquire an expectation value in the 
mean field approximation. This $(Z_2\times Z_2)^2$ is not 
a symmetry of QCD. As a result, we expect that ${\cal O}_{1,2}$
will develop an expectation value once higher order perturbative
corrections are included. Another way to look at this phenomenon
is the observation that the color-flavor locked order parameter
couples left and right handed fields only indirectly, through
the vector-like character of the gauge symmetry. A non-vanishing
expectation value for an operator of the form $(\bar LL)(\bar RR)$
only arises at higher order in perturbation theory. 

  Since chiral symmetry is broken, we also expect that the 
standard chiral order parameter $\langle \bar\psi\psi\rangle$
acquires an expectation value. It was noted in \cite{ARW_98b,RSSV_99} 
that the color-flavor locked phase has an approximate $(Z_2)_L\times
(Z_2)_R$ symmetry. If this symmetry were exact, then the quark
condensate would be zero. In QCD instantons break $(Z_2)_L\times
(Z_2)_R$ to $(Z_2)_V$ and lead to a non-vanishing quark condensate. 
In \cite{RSSV_99} we calculated the quark condensate at moderate
densities, assuming that instantons dominate not only the 
quark condensate, but also the superfluid gap. At very large
density, instantons are suppressed and the gap equation is 
dominated by perturbative effects. The results of \cite{RSSV_99}
are easily generalized to this case.

 In QCD with three flavors, the instanton induced interaction
between quarks is given by \cite{tHo_76,SVZ_80c,SS_98}
\bea
\label{l_nf3}
{\cal L} &=& G \frac{1}{6N_c(N_c^2-1)}
 \epsilon_{f_1f_2f_3}\epsilon_{g_1g_2g_3}
 \left( \frac{2N_c+1}{2N_c+4}
  (\bar\psi_{L,f_1} \psi_{R,g_1})
  (\bar\psi_{L,f_2} \psi_{R,g_2})
  (\bar\psi_{L,f_3} \psi_{R,g_3}) \right. \\
& & \left. + \frac{3}{8(N_c+2)}
  (\bar\psi_{L,f_1} \psi_{R,g_1})
  (\bar\psi_{L,f_2} \sigma_{\mu\nu} \psi_{R,g_2})
  (\bar\psi_{L,f_3} \sigma_{\mu\nu}\psi_{R,g_3})
  + ( L \leftrightarrow R ) \right) \nonumber .
\eea
The coupling constant $G$ is determined by a perturbative
calculation of small fluctuations around the classical
instanton solution. The result is
\be
\label{G_inst}
G = \int d\rho \ n(\rho,\mu)  \ (2\pi\rho)^6, 
\ee
with 
\bea 
\label{nrho}
n(\rho,\mu) &=& C_{N} \ \left(\frac{8\pi^2}{g^2}\right)^{2N_c} 
 \rho^{-5}\exp\left[-\frac{8\pi^2}{g(\rho)^2}\right]
 \exp\left[-N_f\rho^2\mu^2\right],\\
C_{N} &=& \frac{1}{(N_c-1)!(N_c-2)!}\,
 0.466\exp(-1.679N_c)1.34^{N_f}. \nonumber
\eea
At zero density, the $\rho$ integral in (\ref{G_inst}) is
divergent at large $\rho$. This is the famous infrared 
problem of the semi-classical approximation in QCD. At 
large chemical potential, however, everything is under 
control and $G$ is reliably determined. We find
\be
\label{G_res} 
 G = \frac{1}{2}\frac{(2\pi)^6}{6N_c(N_c^2-1)}
     C_{N}S_0^{2N_c}\Lambda^{-5} N_f^{-\frac{5+b}{2}}
     \left(\frac{\Lambda}{\mu}\right)^{5+b}
     \Gamma\left(\frac{b+5}{2}\right).
\ee
Here, $\Lambda$ is the QCD scale parameter, $b=\frac{11}{3}
N_c-\frac{2}{3}N_f=9$ is the first coefficient of the QCD 
beta function and $S_0=8\pi^2/g^2$. The result shows that, 
asymptotically, the coupling constant $G$ has a very strong 
power-law suppression $\sim \mu^{-(5+b)}=\mu^{-14}$. 

 Since $G$ is so small, we can treat the effect of instantons
as a perturbation. In the color-flavor locked phase the instanton 
vertex induces a fermion mass term. This can be seen by saturating
four of the external legs of the interaction (\ref{l_nf3})
with the superfluid order parameter (\ref{phi_cfl}). Using 
the results of \cite{RSSV_99}, we find 
\be 
\label{mass_nf3}
 {\cal L} =   ( \bar \psi_{L,i}^a 
    \frac{1}{2}(1+\vec\alpha\cdot\hat{p})\psi_{R,j}^b)
 \left\{  \left( \delta^{ab}\delta_{ij} \right)\frac{18}{5} 
+ \left( \delta^a_{\;i} \delta^b_{\;j} \right)\frac{6}{5} 
  \right\} G \phi_A^2  + (L\leftrightarrow R).
\ee
We note that there are two kinds of fermion mass terms. 
Both the color singlet structure $(\delta^{ab}\delta_{ij})$
and the color octet structure $(\delta^a_{\;i} \delta^b_{\;j})$
are compatible with the residual $SU(3)_V$ symmetry of the 
color-flavor locked phase. From the result (\ref{mass_nf3}), 
we can directly read off the singlet and octet mass terms 
\be
\label{m_0,8}
 m_0 = \frac{18}{5}G\phi_A^2, \hspace{1cm}
 m_8 = \frac{6}{5} G\phi_A^2.
\ee
From these results we can also determine the quark condensate. 
For this purpose we have to know the momentum dependence of
the fermion propagator in the vicinity of the Fermi surface. 
In principle, this is determined by the momentum dependence 
of the $\mu\neq 0$ fermion zero mode of the instanton, see
\cite{RSSV_99}. For simplicity, we assume that the quark
mass has the same momentum dependence as the gap. In this 
case, we find
\be
\label{qbarq}
\langle \bar\psi\psi\rangle = -
 2\left(\frac{\mu^2}{2\pi^2}\right)\frac{3\sqrt{2}\pi}{g}
 \frac{18}{5} G\phi_A^2.
\ee
From the instanton calculation we can also see how to 
construct a chiral order parameter that is non-vanishing
already on the level of the mean field approximation. 
Consider the 8-fermion operator
\be
\label{o8}
{\cal O}_8 = \det(\bar\psi_L\psi_R)\bar\psi_L\psi_R,
\ee
where $\det(\bar\psi_L\psi_R)$ is a short hand expression
for the flavor structure of the instanton vertex (\ref{l_nf3}).
The determinant is invariant under $SU(3)\times SU(3)_R$, so
${\cal O}_8$ transforms like the quark condensate. From the
discussion above it is clear that ${\cal O}_8$ acquires an
expectation value in the color-flavor locked phase. We find
\be
 \langle {\cal O}_8\rangle = 6\phi_A^4.
\ee
There is nothing special about ${\cal O}_8$, other eight-fermion
operators are equally good order parameters. 
 
Finally, it is interesting to obtain a few numerical
estimates. For definiteness, we will consider the chemical
potential $\mu=500$ MeV. Also, we will use $\Lambda_{QCD}
=200$ MeV. Following \cite{PR_99b} we will be optimistic
and use $g=4.2$, which corresponds to the maximum of the
gap as a function of $g$. In this case we get a substantial
gap in QCD with two flavors, $\Delta_0(N_f\!=\!2)=130$ MeV. The 
gap in the $N_f=2$ phase of QCD with three flavors is 
$\Delta_0(N_f\!=\!3)=47$ MeV. This large reduction comes
from the factor $N_f^{-5/2}$ in the perturbative expression
for $\Delta_0$. This factor is a reflection of the $N_f$ 
dependence of the screening mass. The color anti-symmetric and 
symmetric order parameters in the color-flavor locked 
phase are $\Delta_A=38$ MeV and $\Delta_S=1.3$ MeV. The 
condensation energy is $\epsilon=-32\, {\rm MeV}/{\rm fm}^3$. 
The superfluid order parameter is given by $\phi_A=(144\, 
{\rm MeV})^3$. From this we find that the chiral condensate is 
$\langle\bar\psi\psi\rangle =-(35\, {\rm MeV})^3$,
while the expectation value of the four fermion operator is 
significantly bigger $\langle(\bar\psi\psi)^2\rangle
= (194\,{\rm MeV})^6$.

\section{Conclusions}

 In this work we studied the structure of QCD with an arbitrary 
number of flavors at high baryon density. We assumed that
the dominant instability of the quark fermi liquid is towards 
the formation of a pair condensate. In order to study 
the ground state, we introduced a renormalized mean field 
grand canonical potential which determines the condensation 
energy as a function of the color-flavor structure of the 
gap matrix. This potential is not derived from QCD, but we 
expect it to correctly represent the essential dynamics of 
the high density phase. From the analysis of the grand potential 
we find that for any number of flavors greater than two, chiral 
symmetry is broken while a vector-like flavor symmetry remains. 
The most interesting case is QCD with $N_f=3$ flavors. In
this case, the symmetry breaking pattern at high density 
matches the one at low density. The color-flavor locked 
state in $N_f=3$ is also distinguished by the fact that 
it has the biggest condensation energy per flavor, and that
it leaves the largest subset of the original flavor symmetry
intact. 

 In the second part of this work we studied the $N_f=3$
phase in more detail. We calculated the color symmetric
and anti-symmetric order parameters in perturbative QCD.
We showed that $\Delta_S$ is suppressed by one power of
$g$ in the weak coupling limit, and that the color-flavor
locked state is the energetically favored state. We 
calculated $\langle\bar\psi\psi\rangle$ and $\langle
(\bar\psi\psi)^2\rangle$. We explicitly showed that 
the chiral condensate is suppressed because of an 
approximate $Z_2\times Z_2$ symmetry. In general, 
the expectation value of four-fermion operators is
not suppressed, but all gauge invariant four-fermion
chiral order parameters vanish in the mean field
approximation. Non-vanishing gauge invariant chiral 
order parameters can be obtained by considering higher 
dimension operators, semi-classical (instanton) effects, or
higher order perturbative corrections.

 Acknowledgements: I would like to thank Frank Wilczek
for many useful discussion. Part of this work was 
done during the workshop on ''QCD under extreme conditions''
at the Aspen Center of Physics. This work was supported
in part by NSF PHY-9513385 and the Natural Science and
Engineering Council of Canada.

\section{Erratum}

 There is a mistake in equs.~(11) and (12) in which
we describe the symmetry breaking pattern in the case
$N_c=3$, $N_f=4$. The second set of $SU(2)$ generators 
given in equ.~(12) does not generate a symmetry of the 
order parameter (and it does not commute with the 
first set of generators). The correct generators are
\be
\left(\epsilon^{abc}N_{bc}+
    \frac{1}{2}\eta^a_{\mu\nu}M^{\mu\nu}\right), \hspace{0.5cm}
\left(\frac{1}{2}\bar\eta^a_{\mu\nu}M^{\mu\nu}\right).
\ee
Note that the second $SU(2)$ is a pure flavor symmetry. 
As a consequence, the symmetry breaking pattern is 
\be
\label{sym_4p}
 SU(4)_L\times SU(4)_R\times U(1)_V\to SU(2)_V\times SU(2)_V
\times SU(2)_A 
\ee
and the axial $SU(4)_A$ symmetry is not completely broken. 
I thank H.~Murayama for pointing out this error. 

There is a factor 2 mistake in equation (31) for the 
thermodynamic potential. To $o(g^2)$ the thermodynamic 
potential is given by
\bea
\label{f_ng_2p}
\Omega &=& \frac{1}{2}\int\frac{d^4q}{(2\pi)^4}
 \left\{-{\rm tr}\left[S(q)\Sigma(q)\right]
       +{\rm tr}\log\left[S_0^{-1}(q)S(q)\right]\right\}\\
 & & \mbox{}+ 
\frac{1}{4}\int\frac{d^4q}{(2\pi)^4}\frac{d^4k}{(2\pi)^4}
 {\rm tr}\left[S(q)\Gamma_\mu^a S(q+k)\Gamma_\nu^b\right]
 D^{ab}_{\mu\nu}(k),
\eea
where $\Gamma_\mu^a$ is the quark gluon vertex function.
The last term can be simplified using the gap equation. 
We find 
\be
\label{f_ng_2}
\Omega = \frac{1}{2}\int\frac{d^4q}{(2\pi)^4}
 \left\{-\frac{1}{2}{\rm tr}\left[S(q)\Sigma(q)\right]
       +{\rm tr}\log\left[S_0^{-1}(q)S(q)\right]\right\}.
\ee
Note that the first term differs from equ.~(31) by a 
factor 1/2. As a consequence, equ.~(32) should read 
\be
f(\Delta) = \frac{\mu^2}{\pi^2}  \int dp_0 \left\{ 
-\frac{\Delta(p_0)^2}{2\sqrt{p_0^2+\Delta(p_0)^2}}
+\sqrt{p_0^2+\Delta(p_0)^2}-p_0 \right\}.
\ee
and equ.~(33) is replaced by 
\be
f(\Delta) = \frac{\mu^2}{4\pi^2} \Delta_0^2,
\ee
which is identical to the result in BCS theory.

 Finally, equ.~(48) gives an estimate of the 
quark condensate in the CFL phase. The complete 
weak coupling result is
\be 
\langle\bar{\psi}\psi\rangle = 
-2\left(\frac{\mu^2}{2\pi^2}\right) 
 12G\phi_A^2,
\ee
which was obtained in T.~Sch{\"a}fer,
Phys.~Rev.~D65, 094033 (2002).


\newpage\noindent

\begin{table}
\begin{tabular}{|c|c|c|l|c|c|c|c|}
$N_c$ & $N_f$ & $N_{par}$  &  gaps (deg)  &  $\Delta$
    & $-\epsilon/(N_cN_f)$  &   $N_{sym}$  & rank\\ \hline\hline
  3   &   2   &     3      & $\Delta$ (4), 0 (2)
&   $\Delta_0$   & $\epsilon_0$ &         6 & 2 \\  \hline
  3   &   3   &     9      & $\Delta$ (8), $2\Delta$ (1)
&   $0.80\Delta_0$ & $1.27\epsilon_0$  &  8  &  2\\   \hline
  3   &   4   &    18      & $\Delta$ (8), $2\Delta$ (4)
&   $0.63\Delta_0$ & $1.21\epsilon_0$  &  6  &  2\\    \hline
  3   &   5   &    30      & $\Delta$ (5), $2\Delta$ (7), $3\Delta$ (3)
&   $0.43\Delta_0$ & $1.18\epsilon_0$  &  3  &  1\\    \hline
  3   &   6   &    45      & $\Delta$ (16), $2\Delta$ (2)
&   $0.80\Delta_0$ & $1.27\epsilon_0$  &  9  &  3\\
\end{tabular}
\vspace*{0.5cm}
\caption{Spectrum and symmetry properties of the $s$-wave superfluid
state in QCD with $N_c=3$ colors and $N_f$ flavors. $N_{par}
=N_c(N_c-1)N_f(N_f-1)/4$ is the number of totally anti-symmetric
gap parameters. The fourth column gives the relative magnitude 
of the gaps in the fermion spectrum, together with their degeneracy. 
The next two columns give the magnitude of the gap and the 
condensation energy per species in units of the $N_f=2$ values.
These ratios are independent of the coupling in the weak-coupling
limit. The last two columns show the dimension and the the rank of 
the residual symmetry group.}
\end{table}

\end{document}